\documentclass[preprint,authoryear,11pt]{elsarticle}

\usepackage[T1]{fontenc}
\usepackage{amsmath,amssymb}
\usepackage{graphicx}
\usepackage{booktabs}
\usepackage{microtype}
\usepackage{url}
\usepackage[hidelinks]{hyperref}

\usepackage{tikz}
\usetikzlibrary{positioning, arrows.meta, calc}
\definecolor{biocol}{HTML}{C9663F}
\definecolor{biocoldark}{HTML}{8B3F1E}
\definecolor{sslcol}{HTML}{2E5A87}
\definecolor{sslcoldark}{HTML}{1A3D5C}
\definecolor{fusecol}{HTML}{5A6B7D}
\definecolor{fusecoldark}{HTML}{3A4754}

\newcommand{\modelname}{MOSAIC}
\journal{Speech Communication}

\begin{document}

\begin{frontmatter}

\title{When and why do handcrafted cues help self-supervised
anti-spoofing? A causal and faithfulness analysis}

\author{Yugwon Won}
\address{AI Security R\&D Team, RaonSecure Co., Ltd., Seoul, Republic of Korea}
\ead{linky1584@gmail.com}

\begin{abstract}
Most spoofing countermeasures now place a light classifier on top of a
self-supervised (SSL) speech encoder. A growing line of work adds
handcrafted acoustic features to this design and fuses them by
cross-attention, partly because the attention weights appear to explain
which cues the model relies on. Two things about such fusion remain
untested: whether the handcrafted features contribute anything once a
strong SSL encoder is already in place, and whether the attention map
faithfully reflects what the classifier actually uses. We study both
questions with \modelname{}, a single model that handles logical access
(LA) and physical access (PA) attacks, and that projects a 152-dimensional
biophonetic vector into six query tokens attending over thirteen
intermediate layers of WavLM-Large. Rather than pursuing state-of-the-art
accuracy, we treat the model itself as the object of analysis and apply
inference-time interventions that zero out selected components. Removing
the handcrafted branch produces opposite outcomes in the two domains.
Under LA the removal costs nothing and in fact lowers the EER by 0.67
percentage points, which indicates that WavLM alone is sufficient there. Under PA
the same operation raises the EER by 0.76 percentage points, so the branch supplies
replay channel evidence that the SSL encoder does not fully carry. The
$6\times13$ attention map is highly stable under resampling (bootstrap
rank correlation $0.99$), yet its faithfulness varies by domain: layer
attention weights predict the causal importance of each layer under PA
($\rho={+}0.62$) but not under LA ($\rho={-}0.28$). Taken together, the
handcrafted branch and its attention-based explanation are trustworthy
for replay attacks and not for synthetic ones. On standard benchmarks
the unified model is mid-range on LA while clearly ahead on cross-source
deepfakes (6.21\,\% EER on ASVspoof~2021~DF, against 17.0\,\% for AASIST
and 22--25\,\% for the official baselines), and its apparently weak
2021~PA figure turns out to characterize that benchmark rather than this
model. The practical contribution is not a new architecture but a
checking procedure: verify handcrafted cues and attention maps by
intervention before trusting either.
\end{abstract}


\end{frontmatter}

\section{Introduction}
\label{sec:intro}

Spoofing countermeasures for speaker verification have converged on a
fairly standard template. A self-supervised speech encoder such as
WavLM~\citep{chen2022wavlm} or wav2vec~2.0~\citep{baevski2020wav2vec2}
supplies intermediate representations, and a light back-end such as
AASIST~\citep{jung2022aasist} or attentive merging of hidden
states~\citep{pan2024attentivemerging} reads a spoofing decision out of
them. Pre-trained on thousands of hours of speech, these encoders
already carry much of the low-level acoustic evidence that separates
genuine from synthetic speech, and they have pushed the EER below 1\,\%
on the logical access (LA) condition of ASVspoof, that is, on
text-to-speech and voice conversion.

Handcrafted features are frequently added to this template in order to
complement the encoder. Mel and linear cepstra, constant-Q cepstra,
prosody, and phase-derived quantities are combined with SSL features by
concatenation, by gating, and increasingly by
cross-attention~\citep{elkheir2025twoviews}. We let the handcrafted side query the SSL side, for two reasons. First,
handcrafted features encode physical prior knowledge that a
data-driven encoder has no particular reason to have learned. Second,
the attention weights produced during fusion appear to show which cue is
consulted at which layer, which is useful whenever a detector has to
justify its decision.

Both advantages, however, rest on assumptions that conventional
reporting cannot test. Because a fusion system is evaluated by a single
error rate, a handcrafted branch that contributes nothing, or that
actively hurts, is indistinguishable on the page from one that helps,
and no aggregate number can separate the two. Attention maps are
likewise presented as heatmaps and read off in prose, even though a long
line of work in other fields has shown that attention weights need not
agree with the features a model actually depends
on~\citep{jain2019attention, wiegreffe2019attention,
liu2022faithfulness}. Such checks are rarely performed in anti-spoofing,
and design decisions of real consequence (whether to keep a handcrafted
branch, whether to trust an attention map) are therefore made without
evidence.

We treat a fusion detector of this kind as an object of analysis rather
than as something to be improved. The detector is \modelname{}, which
handles LA and physical access (PA, replay) attacks in one model. It
divides a 152-dimensional handcrafted biophonetic vector (glottal-source,
phase, and channel-response features) into six semantic-group query
tokens and lets them attend over thirteen intermediate layers of
WavLM-Large, producing a $6\times13$ cue-by-layer attention matrix. The
architecture is deliberately ordinary; what is new is the analysis
performed on top of it. We ask two questions:
\begin{itemize}
\item[\textbf{Q1.}] Under which attack conditions does the handcrafted
      branch actually change the decision, and by how much?
\item[\textbf{Q2.}] Does the $6\times13$ attention map describe what the
      model relies on, both stably and faithfully?
\end{itemize}

We answer both by intervening on a trained model at inference time. By
intervention we mean an operation in which the experimenter artificially
sets a given component to zero, or otherwise blocks it, and observes how
performance changes; no retraining is involved. For Q1 we zero the
handcrafted branch and measure the change in EER. For Q2 we remove one
WavLM layer at a time from the attention path and ask whether the layers
that receive the most attention are the ones whose removal costs the
most. The outcomes were pronounced in both cases, and they pointed the
same way. Under LA the handcrafted branch was not merely redundant:
removing it improved the EER, and its attention map did not track causal
importance. Under PA the branch was necessary and the attention map was
faithful. The handcrafted stream and its explanation can therefore be
trusted when the attack is a channel phenomenon (replay) and not when it
is a synthesis phenomenon (LA), which is exactly what the physical
rationale for introducing these features predicts.

Our contribution is thus methodological and empirical rather than
architectural.
\begin{enumerate}
\item Using a branch-removal intervention, we show that the contribution
      of a handcrafted stream to an SSL detector is domain-dependent and
      can be negative. It worsens the EER by 0.67 percentage points under
      LA and improves it by 0.76 points under PA (Section~\ref{sec:causal}). No
      single-metric comparison can reveal this.
\item We test the fusion attention map for stability and for
      faithfulness and show that the two come apart. The map is highly
      reproducible, but it is faithful only under PA
      (Section~\ref{sec:faith}).
\item We compare one unified LA+PA model against reproduced and official
      baselines, confirming a large margin on cross-source deepfakes and
      showing that the one condition where performance is poor
      (2021~PA) is difficult for every system, not for this model alone
      (Section~\ref{sec:detect}).
\end{enumerate}

\section{Related work}
\label{sec:related}

\subsection{Self-supervised front-ends for anti-spoofing}
The dominant design places a light classifier on a self-supervised
encoder. wav2vec~2.0~\citep{baevski2020wav2vec2},
HuBERT~\citep{hsu2021hubert}, the cross-lingual
XLS-R~\citep{babu2021xlsr}, and WavLM~\citep{chen2022wavlm} all provide
frame-level representations that transfer well to spoofing detection,
and back-ends such as AASIST~\citep{jung2022aasist} or attentive merging
of hidden states~\citep{pan2024attentivemerging, guo2024wavlm} read
decisions out of them at sub-1\,\% EER on ASVspoof~2019 LA. One finding
recurs across these studies: the discriminative information is
concentrated in particular layers, with the lower and middle blocks that
preserve acoustic and phonetic detail mattering more than the upper
blocks that carry lexical content~\citep{pan2024attentivemerging}. Such
layer analyses, however, remain correlational, since they read off
attention or probing weights. Our layer-removal experiment
(Section~\ref{sec:faith}) is the causal counterpart: it reproduces the
concentration on layers 6--8 and then tests whether that concentration
reflects actual dependence.

\subsection{Handcrafted features and SSL--handcrafted fusion}
Before SSL front-ends, countermeasures were built from cepstral features
with GMM or CNN back-ends. Linear-frequency cepstral coefficients
(LFCC)~\citep{sahidullah2015}, mel-frequency cepstral coefficients, and
constant-Q cepstral coefficients (CQCC)~\citep{todisco2017cqcc} are the
representative examples, and CQCC-GMM and LFCC-GMM remain official
ASVspoof baselines~\citep{asvspoof2019}. Learnable but physically
interpretable front-ends such as SincNet~\citep{ravanelli2018sincnet}
narrowed the gap to hand-designed filters. For replay detection,
linear-frequency cepstra and sub-band energies are preferred because
they expose the frequency response of the recording and playback path.
Source features such as jitter and shimmer differ measurably between
genuine and synthetic speech, and shimmer in particular varies over a
wider range for synthetic speech~\citep{li2023jittershimmer}. Recent
work reintroduces these features alongside SSL representations and fuses
them by concatenation, gating, or cross-attention. El~Kheir et
al.~(\citeyear{elkheir2025twoviews}) are representative: they use SSL
features as queries and spectral features as keys and values, obtaining
a substantial relative EER reduction over an SSL-only baseline. These
studies, however, report a single fusion error rate, and where they do
examine the contribution of each stream they do so by reading weights
inside the model rather than by intervening on it. Neither the
handcrafted contribution nor the map itself is tested causally, and
those are the two gaps we address.

\subsection{Unified LA/PA detection and domain shift}
Most systems specialize in one attack family, being built for LA/DF or
for PA but rarely for both. The two families leave different
traces, synthesis artifacts on one side and channel artifacts on the
other, and a shared latent space tends to collapse onto whichever axis
is easier to separate. Domain-adversarial training~\citep{ganin2016dann}
is one remedy, and we keep the distinction by using two gradient-reversal
discriminators, one per attack family. As evaluation moved from the
controlled recordings of ASVspoof~2019 to the real microphones and rooms
of ASVspoof~2021~\citep{yamagishi2021asvspoof2021}, and then to recent
corpora such as In-the-Wild~\citep{muller2022itw} and
ASVspoof~5~\citep{wang2024asvspoof5}, cross-condition generalization
became the central open problem~\citep{yi2023survey}.
Section~\ref{sec:detect} documents how a single unified model behaves
across these shifts.

\subsection{Interpretability and the faithfulness of attention}
Attention weights are often offered as explanations and just as often
doubted. Jain and Wallace~(\citeyear{jain2019attention}) showed that
attention distributions can be altered without changing predictions;
Wiegreffe and Pinter~(\citeyear{wiegreffe2019attention}) refined that
claim rather than overturning it; and later work formalized the gap
between attention and causal contribution as a faithfulness
test~\citep{liu2022faithfulness}. The lesson is that a stable or
plausible attention pattern is not by itself evidence of dependence, and
that only intervention can establish it. Anti-spoofing has begun to
value interpretability as well, but it usually visualizes attention or
saliency without testing either. We introduce an intervention-based
faithfulness test and find that its outcome is not uniform across
domains. Attention is faithful under PA and not under LA, which is a
direct warning against reading anti-spoofing attention maps at face
value.

\section{The \modelname{} detector}
\label{sec:method}

\begin{figure*}[t]
\centering
\resizebox{\textwidth}{!}{%
\begin{tikzpicture}[
  x=1cm, y=1cm,
  every node/.style={font=\footnotesize\sffamily},
  arr/.style={-{Latex[length=2.8mm,width=2.2mm]}, line width=1.0pt, draw=gray!50!black},
  biofill/.style={fill=biocol!16, draw=biocoldark!65, line width=0.9pt, rounded corners=6pt},
  sslfill/.style={fill=sslcol!16, draw=sslcoldark!65, line width=0.9pt, rounded corners=6pt},
  fusefill/.style={fill=fusecol!20, draw=fusecoldark!65, line width=0.9pt, rounded corners=6pt},
  headfill/.style={fill=gray!8,  draw=gray!55!black, line width=0.7pt, rounded corners=5pt},
  iofill/.style  ={fill=gray!8,  draw=gray!55!black, line width=0.7pt, rounded corners=5pt},
  biopill/.style={fill=biocol!42, draw=biocoldark, line width=0.6pt, rounded corners=5pt,
                  minimum width=17mm, minimum height=10mm, inner sep=2pt,
                  font=\footnotesize\sffamily\bfseries, align=center},
  qlab/.style={font=\scriptsize\sffamily\itshape, text=biocoldark!85!black},
  grouplab/.style={font=\small\sffamily\bfseries, text=black!80},
]
\draw[iofill] (0, 0.7) rectangle (2.6, 4.3);
\node[font=\small\sffamily\bfseries] at (1.3, 3.8) {Waveform};
\draw[gray!55!black, line width=0.8pt, smooth, samples=80, domain=0.3:2.3]
  plot (\x, {2.5 + 0.27*sin(deg(22*\x)) + 0.14*sin(deg(48*\x+1.2))});
\node[font=\small\sffamily] at (1.3, 1.15) {$x \in \mathbb{R}^{T}$};
\node[grouplab, text=biocoldark, anchor=west] at (3.4, 5.55)
  {Biophonetic Branch (152-dim $\to$ 6 query tokens)};
\draw[biofill] (3.4, 2.95) rectangle (16.2, 5.30);
\foreach \i/\name/\qn in {0/{Praat}/1, 1/{Phase}/2, 2/{LFCC-$\mu$}/3,
                          3/{LFCC-$\sigma$}/4, 4/{Sub-$\mu$}/5, 5/{Sub-$\sigma$}/6}{
  \pgfmathsetmacro{\cx}{4.55 + \i*2.05}
  \node[biopill] at (\cx, 4.35) {\name};
  \node[qlab] at (\cx, 3.35) {$q_{\qn}$};
}
\node[grouplab, text=sslcoldark, anchor=west] at (3.4, -0.55)
  {Self-Supervised Branch (WavLM-Large, layers 6--18)};
\draw[sslfill] (3.4, -0.05) rectangle (16.2, 2.10);
\foreach \i/\hex in {0/2E5A87, 1/365E8A, 2/3F628D, 3/486691, 4/516A94,
                     5/5A6E97, 6/63729A, 7/6C779D, 8/757BA0, 9/7E7FA3,
                     10/8783A6, 11/9087A9, 12/998BAC}{
  \pgfmathsetmacro{\cx}{4.05 + \i*0.55}
  \definecolor{lyc}{HTML}{\hex}
  \draw[fill=lyc, draw=lyc!60!black, line width=0.4pt, rounded corners=1.5pt]
    (\cx, 0.55) rectangle (\cx+0.42, 1.65);
}
\node[font=\scriptsize\sffamily, text=black!80] at (4.26, 0.30) {$L_{6}$};
\node[font=\scriptsize\sffamily, text=black!80] at (7.00, 0.30) {$\cdots$};
\node[font=\scriptsize\sffamily, text=black!80] at (10.30, 0.30) {$L_{18}$};
\draw[arr, sslcoldark, line width=1.1pt] (10.95, 1.10) -- (11.55, 1.10);
\node[font=\footnotesize\sffamily, text=sslcoldark, align=center] at (12.55, 1.10)
  {mean$+$std\\pool};
\node[font=\small\sffamily] at (14.25, 1.10) {$k_{1}\cdots k_{13}$};
\draw[arr] (2.6, 3.0) -- (3.0, 3.0) -- (3.0, 4.10) -- (3.4, 4.10);
\draw[arr] (2.6, 1.5) -- (3.0, 1.5) -- (3.0, 1.05) -- (3.4, 1.05);
\draw[fusefill] (16.95, -0.05) rectangle (22.55, 5.30);
\node[font=\small\sffamily\bfseries, text=fusecoldark, align=center]
  at (19.75, 4.85) {Multi-Token Cross-Attention};
\def\gx{17.55}\def\gy{0.85}\def\gw{4.60}\def\gh{3.10}
\fill[white, rounded corners=2pt] (\gx, \gy) rectangle (\gx+\gw, \gy+\gh);
\draw[fusecoldark!55, line width=0.4pt, rounded corners=2pt]
  (\gx, \gy) rectangle (\gx+\gw, \gy+\gh);
\pgfmathsetmacro\cellw{\gw/13}
\pgfmathsetmacro\cellh{\gh/6}
\fill[fusecol!50, opacity=0.22] (\gx, \gy) rectangle (\gx+3*\cellw, \gy+\gh);
\fill[fusecol!50, opacity=0.11] (\gx+3*\cellw, \gy) rectangle (\gx+7*\cellw, \gy+\gh);
\fill[fusecol!50, opacity=0.05] (\gx+7*\cellw, \gy) rectangle (\gx+\gw, \gy+\gh);
\foreach \c in {1,...,12}{
  \pgfmathsetmacro\xx{\gx + \c*\cellw}
  \draw[fusecoldark!22, line width=0.18pt] (\xx, \gy) -- (\xx, \gy+\gh);}
\foreach \r in {1,...,5}{
  \pgfmathsetmacro\yy{\gy + \r*\cellh}
  \draw[fusecoldark!22, line width=0.18pt] (\gx, \yy) -- (\gx+\gw, \yy);}
\fill[fusecoldark, opacity=0.92]
  (\gx+0*\cellw, \gy+5*\cellh) rectangle (\gx+1*\cellw, \gy+6*\cellh);
\fill[fusecoldark, opacity=0.70]
  (\gx+1*\cellw, \gy+2*\cellh) rectangle (\gx+2*\cellw, \gy+3*\cellh);
\fill[fusecoldark, opacity=0.70]
  (\gx+1*\cellw, \gy+1*\cellh) rectangle (\gx+2*\cellw, \gy+2*\cellh);
\node[font=\scriptsize\sffamily, text=fusecoldark!90, rotate=90]
  at (\gx-0.32, \gy+\gh/2) {6 queries};
\node[font=\scriptsize\sffamily, text=fusecoldark!90]
  at (\gx+\gw/2, \gy-0.32) {13 SSL keys};
\node[font=\scriptsize\sffamily, text=fusecoldark!90]
  at (\gx+\gw/2, \gy+\gh+0.30) {$A\!\in\![0,1]^{6\times13}$~~(4 heads, $d{=}128$)};
\draw[arr] (15.50, 4.10) -- (16.95, 3.70);
\draw[arr] (15.20, 1.10) -- (16.95, 1.40);
\foreach \i/\name/\sub in {0/{Spoof}/{focal}, 1/{LA-DANN}/{A01--A06},
                           2/{PA-DANN}/{AA--CC}, 3/{VAE}/{bona-fide}}{
  \pgfmathsetmacro{\cy}{4.65 - \i*1.35}
  \draw[headfill] (23.30, \cy-0.50) rectangle (26.40, \cy+0.50);
  \node[font=\small\sffamily\bfseries, anchor=center] at (24.85, \cy+0.18) {\name};
  \node[font=\scriptsize\sffamily, text=gray!55!black, anchor=center] at (24.85, \cy-0.22) {\sub};
  \draw[arr] (22.55, 2.60) -- (23.30, \cy);}
\end{tikzpicture}%
}
\caption{The \modelname{} architecture. The handcrafted biophonetic
branch (top) turns the waveform into a 152-dimensional vector split
into six semantic query tokens $q_{1\dots6}$, while the WavLM-Large
branch (bottom) supplies thirteen key/value tokens $k_{1\dots13}$ by
mean--std pooling of layers 6--18. Multi-token cross-attention produces
the $6\times13$ map $A$ together with a fused representation, which
feeds four heads: spoof classification, two domain-adversarial
discriminators, and a bona-fide-only VAE.}
\label{fig:arch}
\end{figure*}
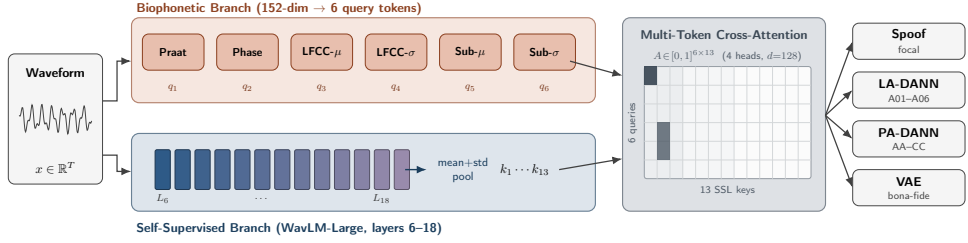

\modelname{} consists of two branches that meet in a cross-attention
block (Figure~\ref{fig:arch}). We describe each branch, the fusion, and
the two intervention operators on which the analysis depends.

\subsection{Biophonetic features}
\label{sec:bio}
By \emph{biophonetic} we mean a set of handcrafted features selected for
their relation to the physics of speech production and of the
transmission channel, rather than for spectral coverage. The vector
$\mathbf{b}\in\mathbb{R}^{152}$ has four groups. \emph{Glottal-source
cues} (9 dimensions) comprise jitter, shimmer, harmonics-to-noise ratio,
voicing ratio, and formant-transition statistics, and reflect the vocal
fold behavior that a vocoder has to approximate. \emph{Phase features}
(7 dimensions) summarize the STFT phase spectrum through phase entropy
and instantaneous-frequency and group-delay statistics, and are
sensitive to the excitation regularity that synthesis tends to smooth
away. \emph{Linear-frequency cepstra} (120 dimensions: 60 temporal means
and 60 standard deviations) and \emph{sub-band energies} (16 dimensions
over 8 bands) characterize the recording and playback channel. The
frequency axis is kept linear rather than mel-warped because the
responses relevant to replay lie in the high band that mel warping
compresses. Concretely, jitter is the mean absolute difference between
consecutive glottal periods $T_i$, normalized by their mean,
\begin{equation}
\mathrm{jitter}=\frac{\frac{1}{N-1}\sum_{i=1}^{N-1}\lvert T_i-T_{i+1}\rvert}
{\frac{1}{N}\sum_{i=1}^{N}T_i};
\end{equation}
group delay is the negative derivative of the unwrapped STFT phase
$\phi(\omega)$, that is
$\tau_g(\omega)=-\partial\phi(\omega)/\partial\omega$; and each
linear-frequency cepstral coefficient is a discrete cosine transform of
the log energies $\log E_m$ of $M$ linear filters,
\begin{equation}
c_k=\sum_{m=1}^{M}\log E_m\,\cos\!\Big[\tfrac{\pi k}{M}\big(m-\tfrac12\big)\Big],
\qquad k=1,\dots,60,
\end{equation}
whose per-frame means and standard deviations form the 120-dimensional
LFCC group. None of these features is new. Jitter and
shimmer~\citep{li2023jittershimmer}, LFCC for
replay~\citep{cmr2025}, and phase features for synthesis are all
established. The term ``biophonetic'' serves to group them into
production-related (glottal, phase) and channel-related (LFCC,
sub-band) sets, a distinction that the causal analysis below makes
concrete.

\subsection{The WavLM branch}
The waveform passes through a frozen WavLM-Large encoder with all hidden
states exposed. For each layer $\ell\in\{6,\dots,18\}$ the frame
sequence is pooled into the concatenation of its temporal mean and
standard deviation, giving
$\mathbf{s}^{(\ell)}\in\mathbb{R}^{2048}$, and a shared linear
projection maps the thirteen pooled vectors into key/value tokens
$S\in\mathbb{R}^{13\times d_s}$ with $d_s{=}512$. Restricting the range
to layers 6--18 follows the layer analyses cited above, since the
topmost and bottommost layers contribute little to spoofing.

\subsection{Cross-attention fusion}
The handcrafted vector is split into six groups $\mathbf{b}_g$, each
projected into a query token by a group-specific affine map with
LayerNorm and GELU,
\begin{equation}
q_g=\mathrm{GELU}\big(\mathrm{LayerNorm}(W_g\mathbf{b}_g+\mathbf{c}_g)\big)
\in\mathbb{R}^{d_b},\qquad d_b=128 .
\end{equation}
Multi-head cross-attention with $h{=}4$ heads lets the six queries read
the thirteen SSL tokens:
\begin{equation}
\mathrm{head}_i=\mathrm{softmax}\!\left(\frac{(QW_Q^i)(KW_K^i)^\top}
{\sqrt{d_k}}\right)VW_V^i ,
\end{equation}
where $Q=[q_1;\dots;q_6]$ and $K{=}V{=}S$. Writing $O_g\in\mathbb{R}^{d_b}$
for the concatenated head outputs of query group $g$ and
$S_\ell\in\mathbb{R}^{d_s}$ for the $\ell$-th row of $S$, the decision is
taken from the concatenation of the attended queries and the mean SSL
token:
\begin{equation}
z=\big[\,\mathrm{mean}_g(O_g)\ \|\ \mathrm{mean}_\ell(S_\ell)\,\big]
\in\mathbb{R}^{d_b+d_s}.
\end{equation}
The by-product of interest is the per-head attention map
$A_i\in[0,1]^{6\times13}$, whose entry $A_i[g,\ell]$ gives how much
query group $g$ attends to WavLM layer $\ell$. We write $A$ for the
head-averaged map, which is what Section~\ref{sec:faith} tests.

\subsection{Training and the intervention operators}
The model is trained with a focal classification
loss~\citep{lin2017focal}, two domain-adversarial terms with separate
gradient-reversal discriminators that keep the LA (A01--A06) and PA
(AA--CC) attack families from collapsing onto a single axis in the
latent space~\citep{ganin2016dann}, and a bona-fide-only variational
regularizer whose reconstruction error on spoofed input serves as an
auxiliary anomaly signal:
\begin{equation}
\mathcal{L}=\mathcal{L}_{\text{focal}}
+\lambda_{\text{LA}}\mathcal{L}_{\text{DANN}}^{\text{LA}}
+\lambda_{\text{PA}}\mathcal{L}_{\text{DANN}}^{\text{PA}}
+\lambda_{\text{rec}}\mathcal{L}_{\text{rec}}^{\text{bona}}
+\lambda_{\text{KL}}\mathcal{L}_{\text{KL}}^{\text{bona}} .
\end{equation}
Each adversarial term is a cross-entropy fed to a discriminator $D_d$
through a gradient-reversal layer $\mathcal{R}_\lambda$, which is the
identity in the forward pass and returns a sign-flipped, rescaled
gradient in the backward pass,
\begin{equation}
\mathcal{L}_{\text{DANN}}^{d}=\mathrm{CE}\big(D_d(\mathcal{R}_\lambda(z)),\,y_d\big),
\qquad \mathcal{R}_\lambda(z)=z,\ \
\frac{\partial\mathcal{R}_\lambda}{\partial z}=-\lambda I ,
\end{equation}
with $d\in\{\mathrm{LA},\mathrm{PA}\}$. The variational branch encodes
$z$ into $(\mu,\sigma)$, is trained on bona-fide input only, and carries
a reconstruction term and a Gaussian KL term in which
$s=\mathrm{mean}_\ell(S_\ell)$ is the mean SSL token and $\hat{s}$ its
reconstruction,
\begin{equation}
\mathcal{L}_{\text{rec}}^{\text{bona}}=\lVert\hat{s}-s\rVert_2^2,\qquad
\mathcal{L}_{\text{KL}}^{\text{bona}}=-\tfrac12\sum_j\big(1+\log\sigma_j^2
-\mu_j^2-\sigma_j^2\big),
\end{equation}
so that at inference a large reconstruction error marks spoofed input as
off-manifold. The analysis requires two inference-time interventions on
the trained network, neither of which needs retraining:
\begin{description}
\item[Branch removal.] The attended query vector
      $\mathrm{mean}_g(O_g)$ is set to zero so that the decision rests on
      the SSL mean alone. The resulting change in EER is the causal
      contribution of the entire handcrafted branch (Q1).
\item[Layer removal.] One WavLM layer $\ell$ is set to zero in the
      attention keys and values while the SSL mean is left intact, which
      isolates the role of that layer along the \emph{attention path}.
      Correlating the resulting change in EER with $A[\cdot,\ell]$ tests
      whether attention tracks importance (Q2).
\end{description}

\section{Experimental setup}
\label{sec:setup}

\subsection{Data}
\modelname{} is trained jointly on the logical and physical access
training partitions of ASVspoof~2019 and then measured on several
evaluation sets. In-domain evaluation uses 2019 LA and 2019 PA;
domain-shift evaluation uses 2021 LA (codec and transmission change),
2021 DF (cross-source deepfake change), and 2021 PA (real rooms and
microphones). To these we add the ASVspoof~5 evaluation
set~\citep{wang2024asvspoof5} as a more recent cross-corpus condition.
The LA training attacks are A01--A06 and the PA conditions span the nine
device and environment codes AA--CC; these labels also supervise the
adversarial branches. Audio is 16~kHz mono and is used at its original
length.

\subsection{Features}
The biophonetic vector (Section~\ref{sec:bio}) has 152 dimensions: 16
production features (9 glottal-source and 7 phase and group-delay), 120
LFCC statistics (temporal means and standard deviations of 60
coefficients), and 16 sub-band energy statistics (means and standard
deviations over 8 bands). LFCCs are computed with a 25~ms window, a
10~ms hop, a 512-point FFT, and 60 linear filters, and the
glottal-source features follow Praat-style estimators. WavLM-Large is
run frozen, and for each of layers 6--18 the temporal mean and standard
deviation of the 1024-dimensional hidden states are concatenated into a
2048-dimensional vector per layer.

\subsection{Model and training}
The six biophonetic groups are projected into $d_b=128$ query tokens and
the thirteen WavLM layers into $d_s=512$ key/value tokens, with $h=4$
attention heads. The fused vector feeds a spoof head, two
domain-adversarial discriminators for the LA and PA attack families
(hidden size 128), and a bona-fide-only variational branch with a
128-dimensional latent. Only the fusion stack is trained: about 2.0M
parameters on top of a frozen 316M-parameter WavLM. We use Adam with a
learning rate of $3\times10^{-4}$ and weight decay $10^{-4}$, batch size
64, and at most 50 epochs, with a three-epoch linear warmup followed by
cosine annealing, gradient clipping at $1.0$, and dropout $0.3$. The
loss weights are focal $\gamma=2.0$, adversarial
$\lambda_{\text{LA}}=\lambda_{\text{PA}}=0.1$, and variational
$\lambda_{\text{rec}}=0.05$ and $\lambda_{\text{KL}}=0.01$; codec
augmentation is applied with probability $0.5$. Early stopping is on
validation EER with patience 20, and the seed is fixed at 42. All
reported figures are ROC-based EERs, and every run was performed on the
MPS backend of a single Apple~M4 Mac~mini.

\subsection{Intervention protocol}
The two interventions of Sections~\ref{sec:causal}--\ref{sec:faith} act
on the trained model at inference time and need no retraining. To sweep
all thirteen layers while still measuring EER changes with adequate
precision, we use a subset of the 2019 evaluation protocols balanced at
2{,}000 bona-fide trials per domain. Branch removal zeros the attended
query vector; layer removal zeros one WavLM layer in the attention keys
and values while preserving the SSL mean branch. The baseline comparison
of Section~\ref{sec:detect}, by contrast, uses the full 2021 evaluation
sets: the four official ASVspoof baselines are scored from the public
challenge score files, and AASIST is run directly on each full set from
the authors' released checkpoint.

\section{Results}
\label{sec:results}

\subsection{Detection performance against baselines}
\label{sec:detect}

Table~\ref{tab:main} compares \modelname{} with the official ASVspoof
baselines, with a reproduced AASIST, and with literature figures for
SSL-based fusion systems. Every number reported there is computed on the
full evaluation set of the corresponding track; for 2021~DF, for
instance, this means all 533{,}928 trials. Training used only 2019 LA and
PA, so every column from 2021 onwards is a cross-corpus condition.

\begin{table}[t]
\centering
\caption{EER (\%) on ASVspoof evaluation sets. The 2021 baseline columns
were computed by the authors from the public challenge score files, and
AASIST 2021 was obtained by running the released checkpoint.
Superscript \textsuperscript{a} marks figures reported in the literature~\citep{asvspoof2019,
jung2022aasist, rawnet2}. LFCC-LCNN was introduced only as a 2021
baseline. The ASVspoof~5 column reports \modelname{} and the reproduced AASIST run
on the full track-1 evaluation set (680{,}774 utterances) under a strict
cross-corpus setting; none of the models here was trained on ASVspoof~5.
Superscript \textsuperscript{b} marks figures reported by El~Kheir et
al.~(\citeyear{elkheir2025twoviews}), who likewise trained only on the
2019~LA partition and evaluated on the same full ASVspoof~5 set, though
their SSL backbone (XLSR-53) and training data (LA only) differ from
ours, so the comparison should be read with care. Dashes indicate that
no figure is available. Column minima are in bold.}
\label{tab:main}
\setlength{\tabcolsep}{5pt}
\resizebox{\textwidth}{!}{%
\begin{tabular}{@{}lccccc@{}}
\toprule
System & 2019 LA & 2021 LA & 2021 DF & 2021 PA & ASVspoof~5 \\
\midrule
\multicolumn{6}{@{}l}{\emph{Official baselines and reproduced AASIST}}\\[2pt]
AASIST     & 0.83\textsuperscript{a} & \textbf{8.15} & 17.03 & 38.16 & 35.53 \\
RawNet2    & 5.13\textsuperscript{a} & 9.50 & 22.38 & 48.61 & --- \\
CQCC-GMM   & 9.57\textsuperscript{a} & 15.62 & 25.56 & \textbf{38.07} & --- \\
LFCC-LCNN  & ---  & 9.26 & 23.48 & 44.77 & --- \\
LFCC-GMM   & 8.09\textsuperscript{a} & 19.29 & 25.25 & 39.54 & --- \\
\addlinespace[3pt]
\multicolumn{6}{@{}l}{\emph{SSL-based fusion systems (reported)\textsuperscript{b}}}\\[2pt]
XLSR-53 $+$ AASIST            & \textbf{0.28} & --- & 5.29 & --- & 23.88 \\
\quad $+$ CQCC cross-attention & 0.40 & --- & \textbf{2.71} & --- & 18.08 \\
\midrule
\modelname{} (proposed) & 1.93 & 9.28 & 6.21 & 40.09 & \textbf{12.57} \\
\bottomrule
\end{tabular}%
}
\end{table}

On cross-source deepfakes \modelname{} separates clearly from the
official baseline family. Its 6.21\,\% on 2021~DF is far ahead of the
17.03\,\% of a reproduced AASIST trained on the same 2019 LA data and of
the 22--25\,\% band into which the official baselines fall. A model with
an explicit channel- and source-sensitive feature set therefore
generalizes to unseen synthesis sources better than one trained from a
raw-waveform front-end alone. Comparison with current SSL-based systems
changes the picture, however: models that fuse handcrafted features with
a large SSL front-end reach 2--5\,\% on the same 2021~DF track and are
ahead of \modelname{}~\citep{elkheir2025twoviews}. Table~\ref{tab:main}
is less a list of best results than a record of where the detector on
which the causal analysis rests happens to stand. On 2021~LA the outcome
is unremarkable: AASIST is lowest at 8.15\,\% and \modelname{} at
9.28\,\% is effectively indistinguishable from RawNet2 or LFCC-LCNN. We
explain why in the causal analysis of Section~\ref{sec:causal}. The
40.09\,\% on 2021~PA looks high in isolation but is unexceptional in
context. Every official baseline falls between 38 and 49\,\% on that set,
RawNet2 reaching 48.61\,\%, and even AASIST, trained on LA alone, gives
38.16\,\% when run from the released checkpoint, because the move from
controlled 2019 PA recordings to the real microphones and rooms of 2021
degrades all of these systems severely. What \modelname{} offers here is
therefore not a lower PA error but the fact that one model produces LA
and PA figures at the same time, in contrast to most prior work, which
reports only one attack family.

The ASVspoof~5 column is the hardest and most recent generalization test
we report. On the full track-1 set ($680{,}774$ utterances) of
crowdsourced speech that also includes newer synthesis systems and
adversarial manipulations~\citep{wang2024asvspoof5}, and without having
seen any of it during training, \modelname{} reached an EER of
12.57\,\%. Compared on the same evaluation set under the same
cross-corpus condition, this is on the low side. AASIST, which we ran on
that set ourselves, gives 35.53\,\%, while prior work reports 23.88\,\%
for an SSL-only baseline and 18.08\,\% for the best handcrafted-fusion
configuration~\citep{elkheir2025twoviews}. Placing these figures side by side shows how much the character of the front-end governs
generalization: the error falls at each step from a raw-waveform
front-end to a large SSL encoder and then to a fusion that adds
handcrafted features, with \modelname{} below them. The model that
trailed on 2021~DF is thus ahead on ASVspoof~5. We take this as a sign
that the channel-related representation acquired by training on LA and PA
together pays off across the varied recording conditions of a recent
corpus. The figure also falls where the earlier results would lead one
to expect. It is well above the 6.21\,\% of the 2021 deepfake track,
where the artifacts remain synthetic and are visible to the SSL
front-end, and well below the 40\,\% of real-world replay, whose channel
neither branch has ever seen. That a model trained only on the 2019
corpora degrades gradually rather than collapsing when applied to a
benchmark released years later suggests that the combination of SSL and
channel features transfers to some degree to new attacks. The size of
that degradation also suggests considerable room for improvement in any
approach that includes ASVspoof~5 in training.

\subsection{Fusion ablation}
Table~\ref{tab:ablation} compares fusion strategies through the
out-of-distribution evaluation used during development. Handcrafted
features alone fall far behind at 17.60\,\% EER, which shows that the
biophonetic set is unsuited to serving as a detector on its own. The
SSL-based variants, whether plain concatenation, single-query
cross-attention, or our six-token cross-attention, all fall between 1.3
and 1.6\,\% and are statistically indistinguishable from SSL alone. The
six-token decomposition is therefore no more accurate, to any measurable
degree, than a single query or than SSL alone, which anticipates the
causal results of Section~\ref{sec:causal}: in terms of aggregate
accuracy, the handcrafted branch and its multi-token split add almost
nothing. What the six-token variant does provide is a readable
cue-by-layer attention map, whose value and limits
Sections~\ref{sec:causal}--\ref{sec:faith} quantify.

\begin{table}[t]
\centering
\caption{Fusion ablation on the development-time out-of-distribution
set. The handcrafted-only system fails and the SSL-based fusion variants
are all comparable, so the six-token split does not buy accuracy.}
\label{tab:ablation}
\begin{tabular}{lcc}
\toprule
Configuration & OOD EER (\%) & min-tDCF \\
\midrule
Handcrafted only              & 17.60 & 0.665 \\
SSL only                      & 1.40  & 0.066 \\
Handcrafted $+$ SSL, concat   & 1.31  & 0.095 \\
Single-query cross-attention  & 1.60  & 0.025 \\
Six-token cross-attention (proposed) & 1.58 & 0.042 \\
\bottomrule
\end{tabular}
\end{table}

\subsection{The causal role of the handcrafted branch (RQ1)}
\label{sec:causal}

\begin{figure}[t]
\centering
\includegraphics[width=0.55\linewidth]{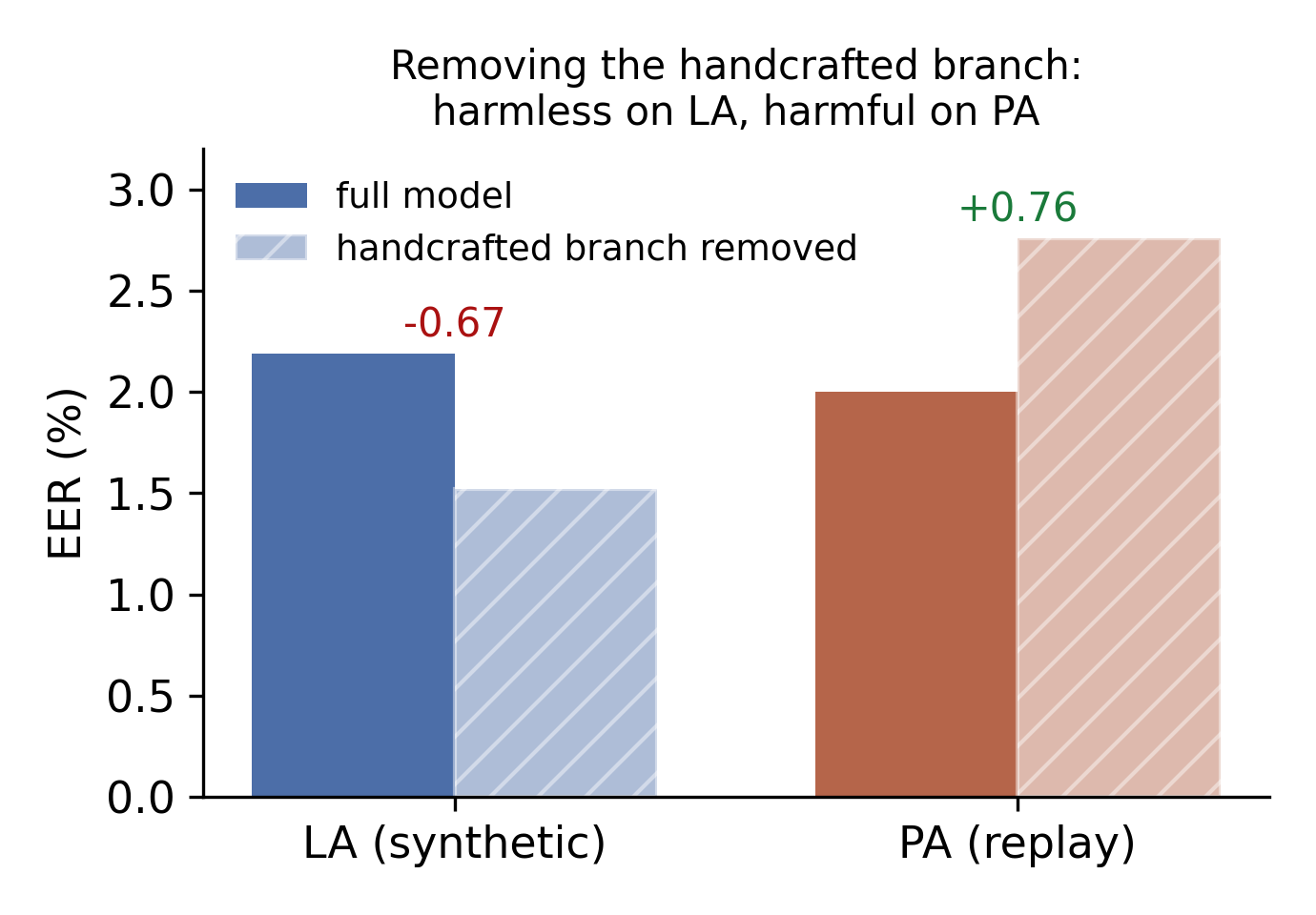}
\caption{Effect of removing the handcrafted branch. Under LA the error
falls, so the branch is a burden; under PA it rises, so the branch is
carrying load.}
\label{fig:causal}
\end{figure}

The most direct way to find out what the handcrafted branch does is to
remove it and observe how the error rate changes
(Table~\ref{tab:causal}, Figure~\ref{fig:causal}). The results were
opposite in the two domains, and that contrast is the point of this
section.

\begin{table}[t]
\centering
\caption{Branch-removal intervention. A positive $\Delta$ means the
handcrafted branch is doing useful work; a negative one means the model
is better off without it. The 95\% intervals are percentile intervals
from 2{,}000 bootstrap resamplings of the evaluation trials; neither
interval contains zero, and the sign of $\Delta$ is consistent across
100\% of the resamplings.}
\label{tab:causal}
\begin{tabular}{lcccc}
\toprule
Domain & Full model & Branch removed & $\Delta$EER & 95\% CI \\
\midrule
LA (synthetic) & 2.19 & 1.52 & $-0.67$ & $[-0.83,-0.39]$ \\
PA (replay)    & 2.00 & 2.76 & $+0.76$ & $[+0.53,+1.07]$ \\
\bottomrule
\end{tabular}
\end{table}

Removing the branch under LA lowers the EER from 2.19\,\% to 1.52\,\%.
The handcrafted queries are thus not merely redundant beside WavLM but a
mild liability, and the model classifies synthetic speech better without
them. This is consistent with the SSL encoder having already become
proficient on LA: the traces left by modern TTS and voice conversion are
visible in the WavLM representation, so adding a low-dimensional
handcrafted summary mostly adds noise to the fused vector.

The aggregate LA effect is not, however, uniform across synthesis
systems. Repeating the branch removal attack by attack, pairing the
bona-fide trials with each of A07--A19 in turn, removal of the
handcrafted stream helps for ten of the thirteen attacks, is neutral for
the two that the model already classifies perfectly (A12 and A13, where
the EER is 0.00\,\% with or without the branch), and does slight harm for
exactly one (A19, ${+}0.50$). More telling is that the harm concentrates
on the attacks that are hardest to judge. The two attacks with the
highest full-model error, A10 (8.95\,\%) and A11 (3.20\,\%), showed the
largest improvements when the branch was removed ($-2.65$ and $-1.10$),
whereas the many attacks that the encoder already handles below 1\,\%
changed only by hundredths of a percent. The handcrafted cues are not a
uniform noise floor, then, but do their damage precisely where the fused
vector is under the greatest discriminative strain, which is what one
would expect of redundancy that the classifier absorbs on easy trials
and fails to absorb on hard ones.

Under PA the same intervention moves the EER the other way, from
2.00\,\% to 2.76\,\%, because here the branch carries information that
the SSL front-end does not supply in sufficient measure. Replay leaves
its trace in the frequency response of the recording and playback
devices and in the characteristics of the room, and it is exactly the
channel-related half of the biophonetic set, the linear-frequency
cepstra and sub-band energies, that makes this trace explicit. The
branch thus does its job in the situation these features were designed
for, and the intervention lets us establish this as a causal statement
rather than a plausible surmise.

In absolute terms neither change is large, at roughly 0.7 percentage points in each
direction, so it is worth being explicit about which conclusion is
robust and which is not. Percentile intervals from 2{,}000 bootstrap
resamplings of the evaluation trials give $[-0.83,-0.39]$ for the LA
effect and $[+0.53,+1.07]$ for the PA effect (Table~\ref{tab:causal});
neither contains zero, and the sign of the effect was preserved in all
2{,}000 resamplings. Our claim therefore rests on the sign of the effect
rather than its magnitude. Even where the handcrafted branch contributes
little to aggregate accuracy, whether that small effect is a help or a
harm divides cleanly along attack families.

The methodological point is that none of these differences would have
surfaced had we reported a single fusion error rate. Presented as one
number, the fusion in \modelname{} looks uniformly competitive across
domains, while intervention reveals that it is counterproductive in one
domain and indispensable in the other. Anyone weighing the adoption of a
handcrafted stream would do well to run this check on their own data
before paying its cost.

\subsection{Stability and faithfulness of the attention map (RQ2)}
\label{sec:faith}

The $6\times13$ attention map is the part of \modelname{} that invites
an interpretability claim. We therefore examine it as two separate
properties, how stable it is and how faithful it is.

\begin{figure}[t]
\centering
\includegraphics[width=0.7\linewidth]{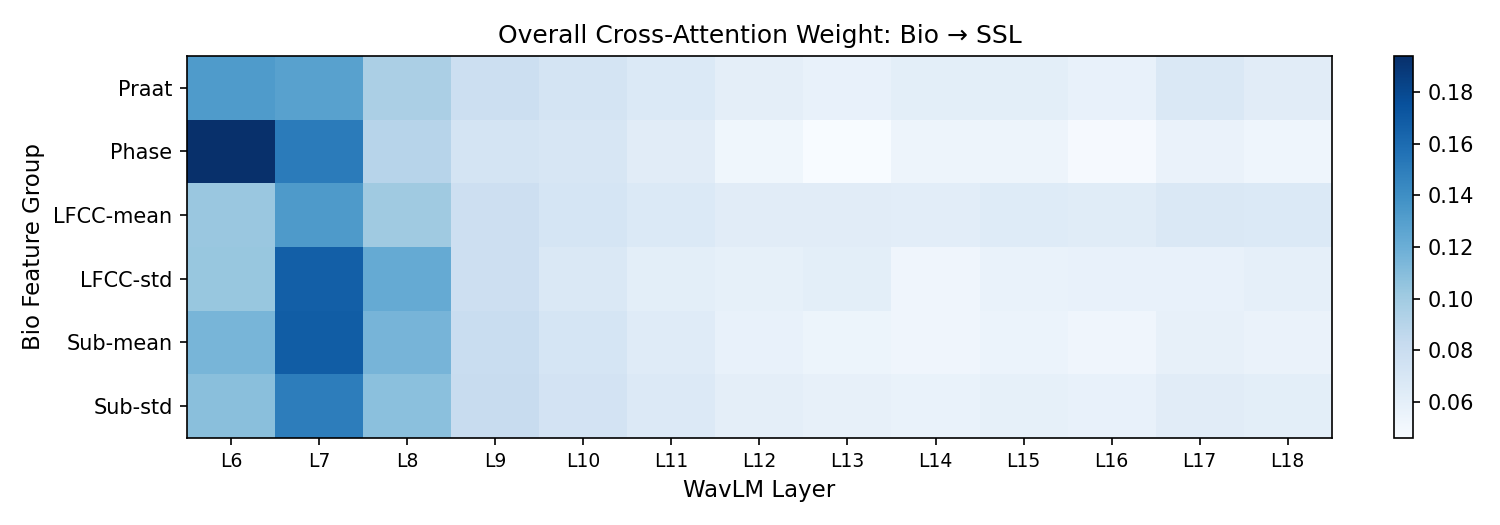}
\caption{Mean $6\times13$ cross-attention map (queries $\times$ WavLM
layers). Attention concentrates on the lower and middle layers and is
highly reproducible under resampling, but stability alone does not make
it faithful (Figure~\ref{fig:faith}).}
\label{fig:attn}
\end{figure}

\paragraph{Stability} We extracted attention over all evaluation trials
and bootstrapped the mean map 200 times. The map barely moves under
resampling: the mean per-cell coefficient of variation is $0.014$, the
mean Spearman correlation between bootstrap maps is $0.993$, splitting
the trials into two disjoint halves still gives $\rho=0.999$, and the
most attended cell stayed the same in all 200 resamplings. The pattern
is consistent with the layer analyses discussed earlier, since attention
concentrates on WavLM layers 6--8 and thins out toward the upper layers
(Figure~\ref{fig:attn}). If stability were sufficient, the discussion
could stop here and the map could be called an explanation.

\begin{figure}[t]
\centering
\includegraphics[width=0.85\linewidth]{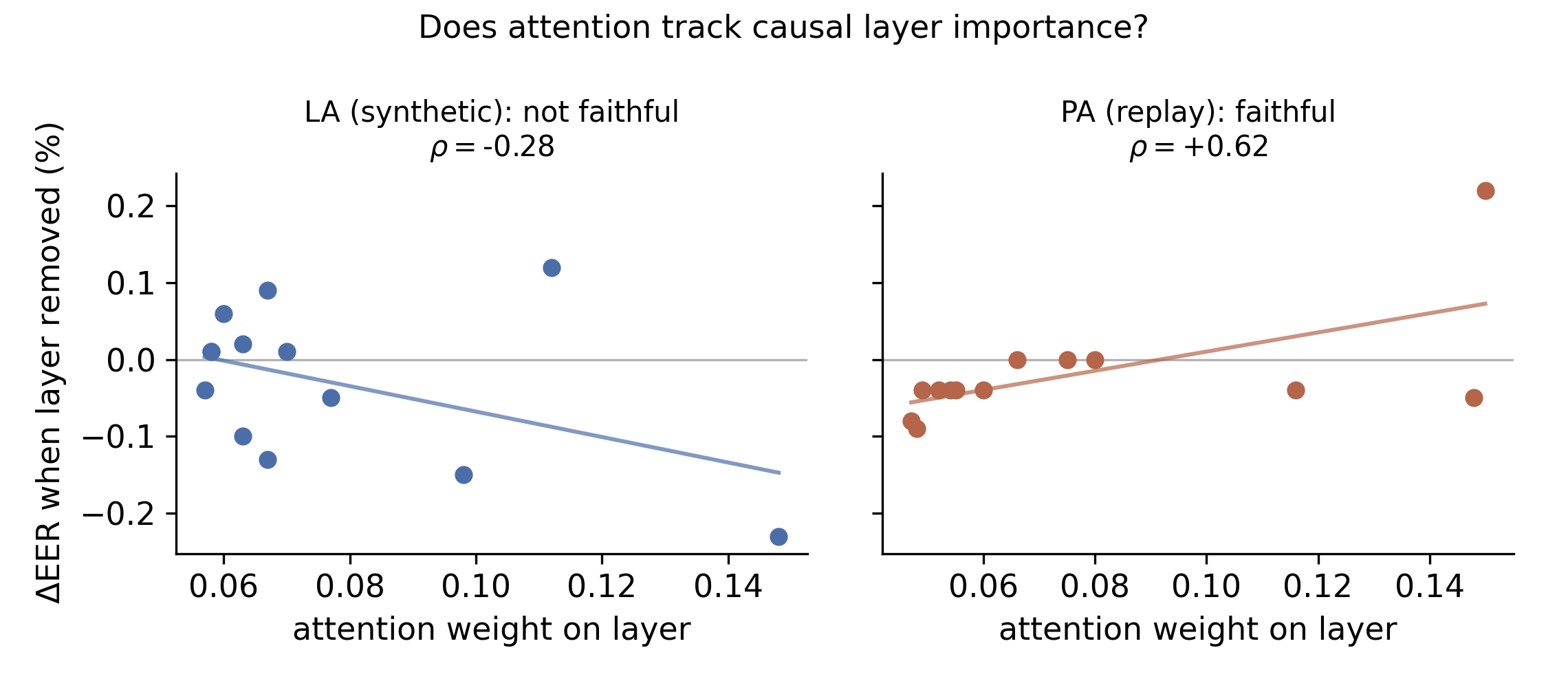}
\caption{Faithfulness test. Each point is one WavLM layer, plotting its
attention weight (x) against the EER increase caused by removing it from
the attention path (y). The two are positively related under PA
(attention is faithful) and not under LA.}
\label{fig:faith}
\end{figure}

\paragraph{Faithfulness} This is a separate matter from stability.
Testing it requires removing WavLM layers one at a time
from the attention path, measuring how far the error rate rises, and
asking whether the layers the model attends to most are the ones whose
removal costs most (Figure~\ref{fig:faith}, Table~\ref{tab:faith}).
Under PA the two quantities move together. The correlation between a
layer's attention weight and the performance loss caused by removing it
reaches $\rho={+}0.62$ ($n=13$, $p=0.03$), showing that heavily attended
layers are indeed the ones holding up performance, so in this case the
map may be read as an explanation of the decision. Under LA the same
correlation is only $\rho={-}0.28$ ($p=0.35$) and is not statistically
distinguishable from zero. With just thirteen layers the negative sign
cannot be read as evidence of anti-correlation, but it is clear that no
grounds for treating the attention as faithful have been established.
The LA attention map, stable and sharply defined as it looks, thus says
nothing about what the LA classifier depends on.

\begin{table}[t]
\centering
\caption{Summary of the two tests. The handcrafted branch and its
attention map are trustworthy under PA and not under LA. Faithfulness is
the Spearman $\rho$ between attention weight and EER increase on
removal, over the thirteen WavLM layers; $p$ is two-sided.}
\label{tab:faith}
\resizebox{\textwidth}{!}{%
\begin{tabular}{lcc}
\toprule
Domain & Branch contribution ($\Delta$EER) & Layer faithfulness ($\rho$) \\
\midrule
LA (synthetic) & $-0.67$ (redundant) & $-0.28$ ($p=0.35$, no evidence) \\
PA (replay)    & $+0.76$ (necessary) & $+0.62$ ($p=0.03$, faithful) \\
\bottomrule
\end{tabular}%
}
\end{table}

The two tests point to the same conclusion. The handcrafted branch
matters under PA and not under LA (Q1), and the attention map is
faithful under PA and not under LA (Q2). Where the branch genuinely
contributes, its attention-based explanation can be trusted; where it
does not, a map that looks like an explanation is merely stable, a
by-product of computation unrelated to the decision. Faithfulness is
thus not a property of the architecture but depends on whether the thing
being explained is actually in use.

\section{Discussion}
\label{sec:disc}

\subsection{The domain-dependent value of handcrafted cues}
Handcrafted cepstral and source features were the foundation of spoofing
detection before SSL front-ends appeared. LFCC and CQCC with GMM
back-ends established the early baselines~\citep{sahidullah2015,
todisco2017cqcc}, jitter and shimmer were identified as cues separating
genuine from synthetic speech~\citep{li2023jittershimmer}, and channel
magnitude response features were devised for replay
detection~\citep{cmr2025}. Recent fusion work reintroduces these
features alongside SSL features and reports aggregate
gains~\citep{elkheir2025twoviews}. Our causal ablation provides grounds
for reading that prior work more precisely: the handcrafted branch is
not always useful but is indispensable for physical access while
redundant for logical access, where it in fact degrades performance
slightly (Section~\ref{sec:causal}). The asymmetry is what one would
expect given the physical traces the two attacks leave in the signal.
Replay leaves a recording and playback channel signature that LFCC and
sub-band energies expose but that a WavLM trained on clean speech does
not represent well, whereas the artifacts of modern TTS and voice
conversion are already visible in the SSL features. Seen this way, it
also becomes clear why replay-specific methods such as
LFCC-CMR~\citep{cmr2025} remain competitive while generic
``add handcrafted features'' approaches yield only marginal gains under
LA: the gain is real but confined to the channel domain. The received
idea that one may fuse anything and let training sort it out therefore
carries a cost. Under LA that approach moves the error rate in the wrong
direction, and without an intervention that removes the branch the fact
does not come to light.

The value of the handcrafted branch splits along domains for reasons
that can be set out at two levels, physical and representational, and it
is the latter that aggregate figures conceal. Physically, a replayed
signal retains the characteristics acquired in passing through recording
and playback devices: the frequency responses of microphone and speaker,
the reverberation of the room, and the timbral coloration added by being
recorded a second time. Among the biophonetic features used here, the
source-related ones, namely jitter, shimmer, and the low-order cepstra,
are sensitive to this coloration, which also connects to the breathiness
and roughness that listeners draw on when judging whether a voice sounds
natural~\citep{li2023jittershimmer}. Representationally, WavLM is
pre-trained to predict masked speech content, and this objective favors
representations that are unaffected by the recording channel, since the
same utterance recorded through different microphones ought to map to
nearly the same features~\citep{chen2022wavlm}. The channel traces that
betray replay are therefore information the encoder learns to suppress,
and making them explicit in cepstral form supplies what the SSL features
no longer carry. Logical access is the opposite case: the temporal and
harmonic regularities that neural vocoders still violate are structure
that a masked-prediction model must represent in order to succeed, so
synthetic artifacts are already separable within the SSL embedding. A
low-dimensional handcrafted summary then merely re-encodes information
that is already present, and adds a little interference in the process.

Seen in this light the finding is not confined to our model but bears on
an assumption implicit in fusion research generally. The dominant
approach is additive, attaching handcrafted features to a strong SSL
front-end and reporting aggregate gains across evaluation sets, and it
rests on the premise that more features can do no harm. Contribution
analyses are not entirely absent: El~Kheir et
al.~(\citeyear{elkheir2025twoviews}) read learned gating weights and
report that the spectral branch accounts for roughly 20\,\% of the fused
representation. That, however, is a correlational analysis of weights
inside the model, not an intervention asking how the error rate changes
when the branch is blocked, and Section~\ref{sec:faith} is precisely a
demonstration that reading weights cannot establish contribution. Our LA
result is then a counterexample to the premise. When combined with an
encoder that already represents the structure needed for
discrimination, an added stream is not neutral but a source of
redundancy that the classifier must learn to discount, and until it
does, the error rate moves the wrong way. The lesson is not that
handcrafted features are obsolete but that their value is conditional on
the information not already being present in the encoder, a condition
that recent surveys still take for granted~\citep{zhang2025survey} and
that we hold must be verified by measurement.

\subsection{Cross-source generalization and the case for a unified model}
The clearest empirical gain in this work is on cross-source deepfakes,
where \modelname{} reaches 6.21\,\% against 17.03\,\% for a reproduced
AASIST and 22--25\,\% for the official baselines, an EER roughly 2.7 times lower than the
strongest of the official and reproduced baselines. This is consistent
with the layer analyses that motivated the design of SSL-based
detectors, since the signal that discriminates spoofing resides mainly
in the lower and middle transformer layers that preserve acoustic
detail~\citep{pan2024attentivemerging}, and combining those layers with
an explicit channel- and source-related feature set transfers to unseen
synthesis systems better than using the SSL front-end alone. Equally important is that the figure comes from a single model that also
produces PA results, in contrast to most
prior work, which specializes in one attack
family~\citep{jung2022aasist, pan2024attentivemerging}. The dual
discriminator adversarial training~\citep{ganin2016dann} was introduced
so that the easier synthesis-versus-replay axis would not dominate the
shared latent space. We did not measure how much this component actually
contributes, so isolating its share through a controlled ablation that
removes the DANN terms remains future work.

What the unified model shows is a fact about representation rather than
a matter of operational convenience. The artifacts left by logical and
physical access attacks are as different as vocoder phase error and room
reverberation, yet the two remain jointly present and still separable
within a single learned representation. If so, the right design
direction for detectors is less a separate specialist model per attack
family than a shared representation from which the cues needed at the
moment can be drawn according to channel conditions. Such a unified
detector is particularly useful in deployment, where it is not known in
advance whether an attack is synthetic or replayed, and the need grows
as conditions move toward the real
world~\citep{yamagishi2021asvspoof2021}.

The one place where this reasoning fails is 2021~PA, where \modelname{}
and every baseline sit together at 38--49\,\% EER. That handcrafted
channel features are causally indispensable on 2019 PA
(Section~\ref{sec:causal}) yet leave the model helpless on 2021 PA looks
contradictory at first. The contradiction dissolves once one sees that
these features capture one particular channel signature rather than a
channel-invariant one. The devices and rooms of 2019 PA are represented
in the training data, so the LFCC and sub-band statistics that describe
them are informative. The recordings of 2021 PA, however, come from
microphones, rooms, and transmission paths that neither branch has ever
encountered, so a cepstral summary tuned to the training channel fails
to transfer to unseen channels just as the SSL features do. The
handcrafted branch, in other words, helps only when the channel
distribution matches training and does not by itself confer robustness
to channel shift. This limitation is shared by dedicated replay
front-ends, and it is why work that models the channel response directly
stresses evaluation across several corpora~\citep{cmr2025}. Narrowing
this gap is therefore less a matter
of finding better handcrafted features than of providing
channel-invariant training signals, through room-response augmentation
or domain adaptation across recording conditions.

\subsection{Faithfulness of attention in anti-spoofing}
The standing of attention as explanation has long been contested outside
speech. Attention was shown to be perturbable without changing
predictions~\citep{jain2019attention}, the claim was subsequently
refined rather than overturned~\citep{wiegreffe2019attention}, and it
was eventually formalized as a faithfulness test measuring the gap
between attention and causal contribution~\citep{liu2022faithfulness}.
Anti-spoofing research has largely bypassed this debate, offering
attention or saliency maps as qualitative evidence of interpretability
without testing them. We introduce an intervention-based faithfulness
test here, and the result is a contribution and a warning at once: the
same attention mechanism that proved faithful under PA ($\rho={+}0.62$,
$p=0.03$) left no grounds for calling it faithful under LA
($\rho={-}0.28$, $p=0.35$). The map is numerically stable in both
cases. The stability that
anti-spoofing research has conventionally taken as reassurance is, in
short, independent of faithfulness.

Why a map can be stable and yet unfaithful is worth setting out
mechanistically, since the two properties are easily confused.
Cross-attention is essentially a routing mechanism: its weights
determine how information is mixed as it enters the fused
representation, not how much each source contributes to the final
decision. What determines the size of the contribution is the
downstream classifier that reads the fused vector. Absent any device
that ties the two together, a highly reproducible attention pattern may
arise from the geometry of queries and keys while the information it
routes contributes nothing to the decision. LA is such a case: the
biophonetic queries consistently draw attention, yet removing the branch
lowers the error, so the routing is stable while its contribution is
near zero. Stability, on this view, indicates only that the routing is
well conditioned, not that it is relevant to the decision; relevance can
be established only by blocking the routed information and observing the
change in output. The same gap has begun to appear in audio deepfake
detection from the standpoint of attribution: post-hoc explanations
obtained from a handful of utterances do not necessarily hold up when
evaluated on large datasets~\citep{grinberg2025xai}, a different route
to the same warning that our interventions reach directly. An
attention heatmap should therefore not be read as an explanation of a
decision until intervention has confirmed that the model actually
depends on what it attends to. The warning carries more weight the more
a detector is expected to justify its own judgments in forensic
settings.

\subsection{A protocol for fusion design and reporting}
Taken together, the two interventions form a lightweight protocol
applicable to any fusion-based countermeasure. Before crediting a
handcrafted stream with a contribution, remove it and examine the change
in error rate broken down by attack family rather than in aggregate.
Before presenting an attention map as an explanation, remove the
components it emphasizes and check whether the decision actually
changes. Both tests operate on an already trained model, so they need no
retraining and complete in a single evaluation pass. We accordingly
propose that these results be reported alongside the usual aggregate
EER, much as ablations and significance tests are already standard in
other fields, since only then do statements such as ``fusion helps'' or
``the model attends to a particular cue'' pass from assertion to
measurement. More broadly, this is a matter of evaluation practice
itself. A fusion error rate shows only that a system works, not which of
its parts actually contributed. So long as countermeasures are compared
by EER alone, there is no way to tell whether the performance came from
the fusion or whether the remaining components compensated for a fusion
that in fact did harm. Reporting causal interventions beside the
headline metric closes this gap at little cost and improves the
reproducibility and interpretability of fusion results. We therefore
regard such reporting as belonging to the basic reporting format for
models of this class rather than as an appendix item.

\subsection{Limitations and future work}
Our detection results extend to the full evaluation set of ASVspoof~5
(Table~\ref{tab:main}), but the causal and faithfulness interventions
themselves were still carried out on a balanced subset of the 2019
corpora, at 2{,}000 bona-fide trials per domain. Extending these
interventions to full evaluation sets, and further to recent real-world
corpora such as In-the-Wild~\citep{muller2022itw} and
ASVspoof~5~\citep{wang2024asvspoof5}, remains future work, since under
such conditions the LA/PA contrast may take a different form or be
blurred by codec and transmission variability. The faithfulness result
is also established for one fusion architecture and one encoder only;
whether the pattern of stable but unfaithful attention recurs with
gating or concatenation fusion, or with other SSL front-ends, has yet to
be checked.

Our intervention operators were deliberately kept simple, and that
simplicity places two constraints on what can be claimed. First, zeroing
an entire WavLM layer in the attention keys and values is a coarse
intervention. Adjacent layers in a large pre-trained encoder tend to
carry overlapping information, so erasing one layer leaves its
neighbors to make up part of its share. Removing a single layer
therefore understates rather than inflates that layer's causal weight.
This works conservatively for the PA result, since a significant
positive correlation ($\rho={+}0.62$) was found even under attenuated
effects. The same cannot be said of the LA result. Whether the absence
of evidence for faithfulness follows from a property of the attention or
from an intervention too blunt to detect the effect cannot be settled
with the present design. An intervention that removes adjacent layers in
blocks, so as to limit redundancy, is the next step toward answering
that question.

Second, branch removal and layer removal act at the granularity of a
whole stream or a whole layer and do not reach the individual
connections linking queries to layers. Severing single cue-to-layer
connections to weigh their share lies beyond what an operator that
replaces values with zero can do, and we have not addressed it here.
Finer interventions, such as removing individual query groups or cutting
single cue-to-layer links, would localize contribution and explanation
more precisely. A stability analysis across training seeds would also
complement the data-resampling stability reported here. A final
practical caveat concerns computational cost. The biophonetic stream is
extracted on CPU through Praat and short-time phase analysis, which adds
a per-utterance cost that the SSL forward pass does not incur, and under
logical access that cost is paid for a stream the model would be better
off without. This is one more reason to measure the contribution of a
handcrafted branch before including it in a deployment pipeline.

\section{Conclusion}
\label{sec:conc}

We have treated a cross-attention detector combining biophonetic and SSL
features as an object of verification rather than of tuning, and put to
it two questions that an error rate cannot answer. The handcrafted
branch turned out to be causally indispensable for the replay attacks of
physical access and a burden for the synthesis attacks of logical
access, and the attention map supplied with it was faithful in the
former case and untrustworthy in the latter. Handling LA and PA in a
single model was itself the precondition for this analysis: had the two
attack families been treated by separate specialist models, the fact
that the contribution of the handcrafted branch reverses across domains
would have been buried in differences of architecture and training
conditions. The detector itself is competitive on LA and ahead of the
official baselines on cross-source deepfakes, but the contribution
likely to last lies in the two interventions rather than in the model,
that is, in a way of actually checking whether a handcrafted stream
earns its cost on a given dataset and whether the attention map that
comes with it can be trusted. With fusion detectors proliferating, we
hope these two tests will come to be reported as a matter of course
alongside error rates. Only then will a fusion figure read not as
evidence that its components help but as a starting point for
determining which of them actually do. The code and evaluation scripts
used in this work are available at
\url{https://github.com/YugwonWon/mosaic-anti-spoofing}.

\section*{Declaration of generative AI and AI-assisted technologies in
the writing process}
During the preparation of this work the author used Anthropic Claude in
order to assist with language editing. After using this tool, the author
reviewed and edited the content as needed and takes full responsibility
for the content of the publication.



\begin{thebibliography}{99}

\bibitem[Babu et~al.(2022)]{babu2021xlsr}
Babu, A., Wang, C., Tjandra, A., Lakhotia, K., Xu, Q., Goyal, N.,
\ldots{} Auli, M. (2022). XLS-R: Self-supervised cross-lingual speech
representation learning at scale. In \emph{Proc. Interspeech}.

\bibitem[Baevski et~al.(2020)]{baevski2020wav2vec2}
Baevski, A., Zhou, H., Mohamed, A., \& Auli, M. (2020). wav2vec 2.0: A
framework for self-supervised learning of speech representations. In
\emph{Advances in Neural Information Processing Systems}.

\bibitem[Bekiryaz{\i}c{\i} et~al.(2025)]{cmr2025}
Bekiryaz{\i}c{\i}, \c{S}., Hanil\c{c}i, C., \& Ozcan, N. (2025). Toward
robust replay attack detection in automatic speaker verification: A
study of spectrum estimation and channel magnitude response modeling.
\emph{Computer Speech \& Language}.

\bibitem[Chen et~al.(2022)]{chen2022wavlm}
Chen, S., Wang, C., Chen, Z., Wu, Y., Liu, S., Chen, Z., \ldots{} Wei,
F. (2022). WavLM: Large-scale self-supervised pre-training for full
stack speech processing. \emph{IEEE Journal of Selected Topics in
Signal Processing}, 16(6), 1505--1518.

\bibitem[El~Kheir et~al.(2025)]{elkheir2025twoviews}
El~Kheir, Y., Zhang, X., Zhao, Y., \& others. (2025). Two views, one
truth: Spectral and self-supervised features fusion for robust speech
deepfake detection. In \emph{Proc. IEEE Workshop on Applications of
Signal Processing to Audio and Acoustics (WASPAA)}.

\bibitem[Ganin et~al.(2016)]{ganin2016dann}
Ganin, Y., Ustinova, E., Ajakan, H., Germain, P., Larochelle, H.,
Laviolette, F., Marchand, M., \& Lempitsky, V. (2016).
Domain-adversarial training of neural networks. \emph{Journal of
Machine Learning Research}, 17(59), 1--35.

\bibitem[Grinberg et~al.(2025)]{grinberg2025xai}
Grinberg, P., Kumar, A., Koppisetti, S., \& Bharaj, G. (2025). What does
an audio deepfake detector focus on? A study in the time domain. In
\emph{Proc. IEEE ICASSP}.

\bibitem[Guo et~al.(2024)]{guo2024wavlm}
Guo, Y., Huang, H., Chen, X., Zhao, H., \& Wang, Y. (2024). Audio
deepfake detection with self-supervised WavLM and multi-fusion attentive
classifier. In \emph{Proc. IEEE ICASSP} (pp.~12702--12706).

\bibitem[Hsu et~al.(2021)]{hsu2021hubert}
Hsu, W.-N., Bolte, B., Tsai, Y.-H. H., Lakhotia, K., Salakhutdinov, R.,
\& Mohamed, A. (2021). HuBERT: Self-supervised speech representation
learning by masked prediction of hidden units. \emph{IEEE/ACM
Transactions on Audio, Speech, and Language Processing}, 29,
3451--3460.

\bibitem[Jain \& Wallace(2019)]{jain2019attention}
Jain, S., \& Wallace, B. C. (2019). Attention is not explanation. In
\emph{Proc. NAACL-HLT}.

\bibitem[Jung et~al.(2022)]{jung2022aasist}
Jung, J.-w., Heo, H.-S., Tak, H., Shim, H.-j., Chung, J. S., Lee, B.-J.,
Yu, H.-J., \& Evans, N. (2022). AASIST: Audio anti-spoofing using
integrated spectro-temporal graph attention networks. In \emph{Proc.
IEEE ICASSP}.

\bibitem[Li et~al.(2023)]{li2023jittershimmer}
Li, K., Baird, A., \& Schuller, B. (2023). Contributions of jitter and
shimmer in the voice for fake audio detection. \emph{IEEE Access}, 11.

\bibitem[Lin et~al.(2017)]{lin2017focal}
Lin, T.-Y., Goyal, P., Girshick, R., He, K., \& Doll\'ar, P. (2017).
Focal loss for dense object detection. In \emph{Proc. IEEE
International Conference on Computer Vision (ICCV)}.

\bibitem[Liu et~al.(2022)]{liu2022faithfulness}
Liu, Y., Li, H., Guo, Y., Kong, C., Li, J., \& Wang, S. (2022).
Rethinking attention-model explainability through faithfulness
violation test. In \emph{Proc. ICML}.

\bibitem[M\"uller et~al.(2022)]{muller2022itw}
M\"uller, N. M., Czempin, P., Dieckmann, F., Froghyar, A., \& B\"ottinger,
K. (2022). Does audio deepfake detection generalize? In \emph{Proc.
Interspeech}.

\bibitem[Pan et~al.(2024)]{pan2024attentivemerging}
Pan, Z., Liu, T., Sailor, H. B., \& Wang, Q. (2024). Attentive merging
of hidden embeddings from pre-trained speech models for anti-spoofing.
In \emph{Proc. Interspeech}.

\bibitem[Ravanelli \& Bengio(2018)]{ravanelli2018sincnet}
Ravanelli, M., \& Bengio, Y. (2018). Speaker recognition from raw
waveform with SincNet. In \emph{Proc. IEEE Spoken Language Technology
Workshop (SLT)}.

\bibitem[Sahidullah et~al.(2015)]{sahidullah2015}
Sahidullah, M., Kinnunen, T., \& Hanil\c{c}i, C. (2015). A comparison of
features for synthetic speech detection. In \emph{Proc. Interspeech}.

\bibitem[Tak et~al.(2021)]{rawnet2}
Tak, H., Patino, J., Todisco, M., Nautsch, A., Evans, N., \& Larcher, A.
(2021). End-to-end anti-spoofing with RawNet2. In \emph{Proc. IEEE
ICASSP}.

\bibitem[Todisco et~al.(2017)]{todisco2017cqcc}
Todisco, M., Delgado, H., \& Evans, N. (2017). Constant Q cepstral
coefficients: A spoofing countermeasure for automatic speaker
verification. \emph{Computer Speech \& Language}, 45, 516--535.

\bibitem[Todisco et~al.(2019)]{asvspoof2019}
Todisco, M., Wang, X., Vestman, V., Sahidullah, M., Delgado, H.,
Nautsch, A., Yamagishi, J., Evans, N., Kinnunen, T., \& Lee, K. A.
(2019). ASVspoof 2019: Future horizons in spoofed and fake audio
detection. In \emph{Proc. Interspeech} (pp.~1008--1012).

\bibitem[Wang et~al.(2024)]{wang2024asvspoof5}
Wang, X., Delgado, H., Tak, H., Jung, J.-w., Shim, H.-j., Todisco, M.,
\ldots{} Lee, K. A. (2024). ASVspoof 5: Crowdsourced speech data,
deepfakes, and adversarial attacks at scale. In \emph{Proc. ASVspoof
Workshop}.

\bibitem[Wiegreffe \& Pinter(2019)]{wiegreffe2019attention}
Wiegreffe, S., \& Pinter, Y. (2019). Attention is not not explanation.
In \emph{Proc. EMNLP-IJCNLP}.

\bibitem[Yamagishi et~al.(2021)]{yamagishi2021asvspoof2021}
Yamagishi, J., Wang, X., Todisco, M., Sahidullah, M., Patino, J.,
Nautsch, A., \ldots{} Delgado, H. (2021). ASVspoof 2021: Accelerating
progress in spoofed and deepfake speech detection. In \emph{Proc.
ASVspoof Workshop}.

\bibitem[Yi et~al.(2023)]{yi2023survey}
Yi, J., Wang, C., Tao, J., Zhang, X., Zhang, C. Y., \& Zhao, Y. (2023).
Audio deepfake detection: A survey. \emph{arXiv preprint
arXiv:2308.14970}.

\bibitem[Zhang et~al.(2025)]{zhang2025survey}
Zhang, B., Cui, H., Nguyen, V., \& Whitty, M. (2025). Audio deepfake
detection: What has been achieved and what lies ahead. \emph{Sensors},
25(7), 1989.

\end{thebibliography}
\end{document}